\documentclass[journal=jacsat,manuscript=article]{achemso}

\usepackage{chemformula} 
\usepackage[T1]{fontenc} 
\usepackage{textgreek}
\usepackage{siunitx}

\author{Man Shen}
\affiliation[Osaka University]
{Department of Applied Physics, Osaka University, Suita, Osaka, 565-0871, Japan}
\email{shen.m@ap.eng.osaka-u.ac.jp}
\author{Taiki Inoue}
\affiliation[Osaka University]
{Department of Applied Physics, Osaka University, Suita, Osaka, 565-0871, Japan}
\author{Mengyue Wang}
\affiliation[Osaka University ]
{Department of Applied Physics, Osaka University, Suita, Osaka, 565-0871, Japan}
\author{Yuanjia Liu}
\affiliation[Osaka University]
{Department of Applied Physics, Osaka University, Suita, Osaka, 565-0871, Japan}
\author{Yoshihiro Kobayashi}
\affiliation[Osaka University]
{Department of Applied Physics, Osaka University, Suita, Osaka, 565-0871, Japan}
\email{kobayashi@ap.eng.osaka-u.ac.jp}

\title
{Efficient defect healing of single-walled carbon nanotubes through C\textsubscript{2}H\textsubscript{2}-assisted multiple-cycle treatment with air exposure}


\begin{document}






\begin{abstract}
Defects in single-walled carbon nanotubes (SWCNTs) degrade their mechanical, electrical, and thermal properties, limiting their potential applications. To realize the diverse applications of SWCNTs, it is essential to enhance their crystallinity through effective defect healing. However, traditional thermal treatments typically require temperatures above 1800°C, which can alter the nanotube structure. Previously, defect healing of SWCNTs was achieved at a relatively low temperature of 1100°C, using C\textsubscript{2}H\textsubscript{2} assistance, but the efficiency was limited. In this study, we developed a C\textsubscript{2}H\textsubscript{2}-assisted multiple-cycle process at an even lower temperature of 1000°C combined with air exposure, achieving highly efficient defect healing while preserving the nanotube structure. The combination of multiple-cycle treatment and air exposure between cycles was found to promote defect activation, suppress the formation of amorphous carbon, and enhance the effectiveness of defect healing. Additionally, we successfully healed commercially available bulk-scale SWCNTs (super-growth SWCNTs), noting that their healing behavior differed from lab-grown SWCNTs with smaller diameters synthesized from nanodiamond. The efficient and structure-preserved healing process developed in this study broadens the potential applications of high-quality SWCNTs, including flexible electronics, high-performance composites, and energy storage devices.
\end{abstract}

\section{Introduction}
Carbon nanotubes (CNTs) have attracted significant research interest due to their exceptional and unique properties since their discovery.\cite{iijima1991helical,iijima1993single,nessim2010properties,bai2021mechanical}
Single-walled carbon nanotubes (SWCNTs), in particular, are considered promising materials for a wide range of applications, including sensors\cite{doi:10.1126/science.287.5453.622,yamada2011stretchable}, electronics\cite{tans1998room, sun2011flexible,shulaker2013carbon,liu2020aligned}, and even futuristic concepts like space elevators.\cite{pugno2006strength,pugno2007space,gohardani2014potential}. Theoretical and experimental studies have demonstrated that SWCNTs possess remarkable mechanical properties, such as a high tensile strength of approximately 100 GPa\cite{jhon2016tensile,bai2018carbon,bai2021mechanical} and a Young's modulus of approximately 1 TPa\cite{Krishnan1998YoungsMO,yu2000tensile}. However, even a single defect (defect concentration 2.7×10\textsuperscript{-3}) in the SWCNTs structure can lead to a 0\%-25\% decrease in performance, while three defects (concentration 8.3×10\textsuperscript{-3}) may cause a reduction of up to 50\%.\cite{shi2022influence} Moreover, synthesized SWCNTs often exhibit lower electrical and thermal conductivities than their theoretically predicted values (e.g., 10\textsuperscript{9} A/cm\textsuperscript{2} for electrical\cite{collins2001engineering} and 3500 W/m·K for thermal conductivity\cite{yu2005thermal,pop2006thermal}), primarily due to the presence of structural defects.\cite{dai1996probing,yamamoto2006nonequilibrium} Defects such as adatoms\cite{tsetseris2009adatom}, vacancies\cite{huang2008vacancy}, Stone-wales defects\cite{zhou2003formation} are commonly introduced during the SWCNTs growth process\cite{yuan2012efficient} or during subsequent purification steps\cite{hersam2008progress,barman2010effects}. Despite substantial progress in the synthesis of low-defect SWCNTs\cite{dai1996single,maruyama2002low,moisala2006single,kimura2012mutual,okada2018flame,lin2018microwave,iakovlev2019artificial,wang2022thermal}, defects remain inevitable, particularly during the physical and chemical purification stages. These defects significantly impair the properties of SWCNTs, underscoring the importance of defect healing as a critical step to fully harness their potential in advanced applications.

Thermal treatment is a widely used method for defect healing in various materials and has also been applied to the post treatment of CNTs. For multi-walled carbon nanotubes (MWCNTs) are treated at temperatures between 1200°C and 2000°C in high vacuum or 2000°C and 2800°C in an Ar atmosphere, has been shown to improve crystallinity.\cite{mattia2006effect,jin2007effect,zhao2012structural} Similarly, heating SWCNTs in high-temperature environments, either in a vacuum or under Ar, has been reported to enhance their crystallinity.\cite{matsumoto2015elucidating,lee2022ultrahigh} However, such thermal treatments typically require temperatures above 2000°C, which can lead to coalescence between individual SWCNTs, thus altering their structure.\cite{yudasaka2001diameter,yudasaka2003structure}
Since the properties of SWCNTs are determined by their structure, including chirality and diameter, uncontrolled structural alterations pose significant challenges for their applications.
Additionally, for SWCNTs grown on substrates, carbothermal reactions between the SWCNTs and the substrate cause decomposition of the SWCNTs at high temperatures and pose another significant
challenge.\cite{koc1998synthesis,nemes2010crystallographically} 
Therefore, achieving defect healing at relatively lower temperatures (below 1200°C) is required for practical applications. Density functional theory simulations have demonstrated that carbon feedstocks, such as C\textsubscript{2}H\textsubscript{2} and C\textsubscript{2}H\textsubscript{4}, can heal vacancy defects in SWCNTs structures.\cite{nongnual2011healing}
Experimentally, we previously improved the crystallinity of SWCNTs at 1100°C with the assistance of C\textsubscript{2}H\textsubscript{2}, though the healing efficiency remained limited.\cite{wang2022thermal}
Moreover, the feasibility of healing bulk-scale SWCNTs with C\textsubscript{2}H\textsubscript{2} has not been fully explored.

In this work, we demonstrate that SWCNTs can undergo further healing after an initial treatment, introducing a novel approach, called multiple-cycle defect healing. Most previous healing studies on CNTs have typically focused on single-cycle healing process, without investigating whether further healing is achievable after an initial treatment. Exploring multiple-cycle healing treatments could offer a more efficient approach and deepen our understanding of the healing mechanisms. Multiple-cycle defect healing incorporating air exposure between cycles, facilitates continuous improvement of crystallinity in ND-CNTs with the assistance of C\textsubscript{2}H\textsubscript{2}. Thanks to the relatively low process temperature, changes in the diameters of SWCNTs are effectively suppressed. Our findings offer new insights into the healing effects of C\textsubscript{2}H\textsubscript{2}, the role of air exposure, and the associated defect-healing mechanisms in SWCNTs. Furthermore, we extend the multiple-cycle defect-healing process to bulk-scale commercial SWCNTs to assess its practical applicability and explore the differences in the healing effects of C\textsubscript{2}H\textsubscript{2} and air exposure between the two types of SWCNTs.

\section{Experimental}\label{sec:experimental}
\subsection{Preparation of ND-CNTs and SG-CNTs}
For the healing experiment, we prepared two types of SWCNTs: ND-CNTs, synthesized in-house by chemical vapor deposition (CVD) using nanodaiamond (ND) as a growth seed\cite{takagi2009carbon,homma2009single,wang2022combination,wang2022thermal}, and commercially available super-growth SWCNTs (SG-CNTs)\cite{hata2004water}. ND-CNTs are particularly suitable for defect-healing experiments as they are grown from ND particles and are virtually free of metal nanoparticles, which could induce undesirable reactions at high temperatures. Similarly, the low metal impurity content in SG-CNTs minimizes the impact of metal impurities during the healing process.

ND-CNTs were synthesized using a previously reported method with slight modifications.\cite{takagi2009carbon,homma2009single,wang2022combination,wang2022thermal} Briefly, purified ND particles were deposited onto a clean Si substrate with a 300-nm thick thermal oxide layer as growth seeds. As shown in the temperature diagram (Figure S1), before growth of ND-CNTs, the ND was pretreated with the surface-cleaning process in air for 10 min at 600°C. ND-CNTs were then synthesized using a three-zone furnace set with a temperature gradient of 850°C-785°C-750°C for 60 min under a mixture of 10 sccm C\textsubscript{2}H\textsubscript{2} (2\%)/Ar and 10 sccm H\textsubscript{2} (3\%)/ Ar.
\cite{wang2022thermal,takagi2009carbon,homma2009single,wang2022combination}
Figure 1(a) and Figure 1(b) show the scanning electron microscopy (SEM) and transmission electron microscopy (TEM) images of the as-synthesized ND-CNTs, respectively. The synthesized ND-CNTs are predominantly single-walled, with lengths reaching several micrometers. The diameter of the ND-CNTs shown in Figure 1(b) is 1.4 nm. Analysis of the TEM images demonstrates that the diameter of the ND-CNTs ranges from 0.5 to 2.6 nm, with an average diameter of around 1.2 nm. After performing Raman characterization on the as-synthesized ND-CNTs, as described later, the samples were used in the subsequent healing experiment. 

The SG-CNTs powder (ZEONANO\textcircled{R}SG101), synthesized via super-growth method (water-assisted CVD),\cite{hata2004water} was purchased from Zeon Nano Technology Co., Ltd. Figure 1(c) and Figure 1(d) show the SEM and TEM images of pristine SG-CNTs, respectively. The diameter of the nanotubes in Figure 1(d) is 3.2 nm, while the SG-CNTs generally have diameters ranging from 3 to 5 nm and lengths 100--600 \textmu m.
For the healing process, approximately 5 mg of SG-CNTs were placed in a graphite crucible, and a graphite lid covering three-quarters of the crucible's opening was used to prevent SG-CNTs from being blown away. The crucible containing the SG-CNTs was placed in the furnace for heating and defect healing. 

\begin{figure}[htbp]
  \centering
   \includegraphics[width=12cm]{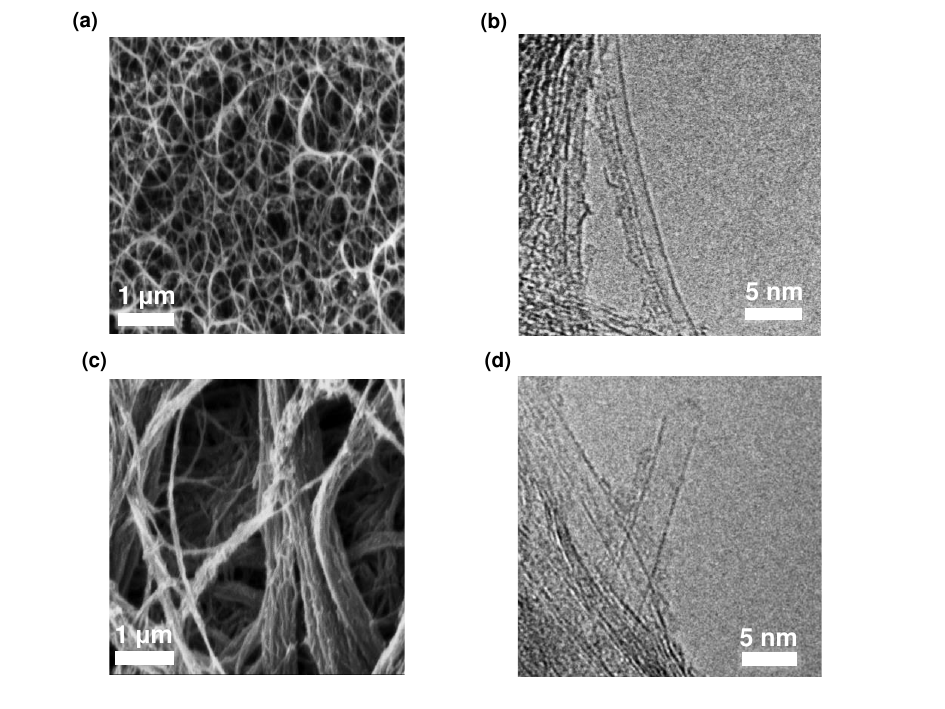}
    \caption{(a)Scanning electron microscopy (SEM) and (b) Transmission electron microscopy (TEM) images of ND-CNTs; (c) SEM and (d) TEM images of SG-CNTs.}\label{figure1}
\end{figure}

\subsection{Multiple-cycle defect healing}
As illustrated in Figure 2 (a), during the multiple-cycle defect-healing process, SWCNTs, either positioned on a substrate or placed within a graphite crucible, were heated in a CVD furnace under an Ar flow of 25~sccm at a pressure of 65~Pa until the temperature reached 1000°C. Upon reaching this temperature, the defect-healing process commenced with the introduction of a gas mixture consisting of 5~sccm~C\textsubscript{2}H\textsubscript{2} (2\%)/Ar and 20~sccm~Ar. The total pressure during the healing process was maintained at 65~Pa, corresponding to a C\textsubscript{2}H\textsubscript{2} partial pressure of 0.26~Pa. The healing process lasted 60~min, after which the furnace was cooled to room temperature (RT, 25°C) under a continuous Ar flow of 20~sccm. Following cooling, the SWCNTs were then removed from the furnace for characterization. Subsequently, the SWCNTs were reintroduced into the furnace for the next healing cycle. Because the samples were exposed to air during characterization, this method is referred as "multiple-cycle defect healing with air exposure".

To assess whether the healing effects were merely due to prolonged annealing, a parallel experiment was conducted using a "single-cycle defect-healing process", as shown in Figure 2(b). In this approach, the SWCNTs were returned to the furnace immediately after synthesis and initial characterization. The furnace was heated under an Ar flow of 25 sccm at a pressure of 65 Pa until reaching 1000°C was reached. As in the multiple-cycle defect healing with air exposure, the healing began with the introduction of a gas mixture of 5 sccm C\textsubscript{2}H\textsubscript{2} (2\%)/Ar and 20 sccm Ar, maintaining the total pressure at 65 Pa and a C\textsubscript{2}H\textsubscript{2} partial pressure of 0.26 Pa. On the other hand, in the single-cycle approach, the healing was conducted continuously for 180 min without interruption. Afterward, the furnace was cooled to RT under a continuous Ar flow of 20 sccm, and the samples were removed for characterization.

To investigate the specific effect of air exposure, another parallel experiment was performed using multiple-cycle healing without air exposure, as illustrated in Figure 2 (c). In this approach, the SWCNTs were returned to the furnace immediately after synthesis and initial characterization. The defect-healing process followed the same conditions as described for the air-exposed method. After each healing cycle, the furnace was cooled to RT, but the samples remained inside the furnace under a continuous Ar flow of 20~sccm throughout the process. For the single-cycle conditions, the samples were exposed to air after the first cycle. For multiple-cycle conditions, the samples underwent consecutive healing cycles without being removed. Upon completion of all healing cycles, the samples were removed from the furnace for final characterization.

\begin{figure}[htbp]
  \centering
   \includegraphics[width=16cm]{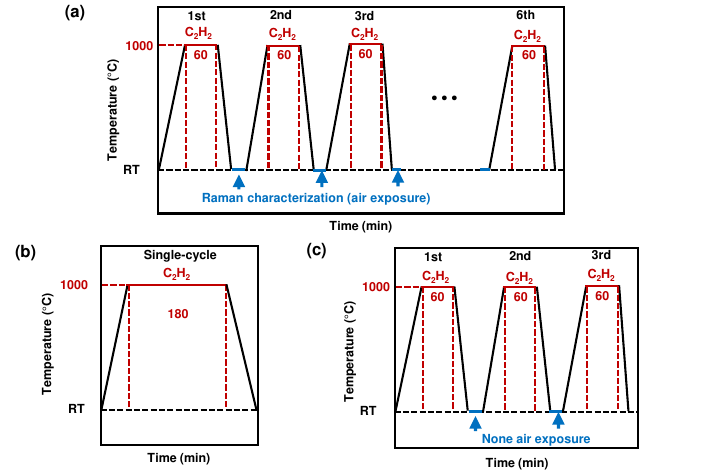}
    \caption{Temperature-time schematic diagram of (a) multiple-cycle defect healing with air exposure; (b) single-cycle defect healing; (c) multiple-cycle defect healing without air exposure. }\label{figure2}
\end{figure}

\subsection{Characterization}
The Raman spectra of the SWCNTs, before and after the defect healing process, were obtained using a Raman spectrometer (LabRAM HR800, HORIBA Jovin Yvon) with an excitation wavelength \textlambda\textsubscript{ex} of 633 nm.
The laser spot size was approximately 0.9 µm, and the laser power at the measurement point was maintained at approximately 9 mW. Each measurement spot was exposed for 1 s over 5 cycles. Raman spectra were collected from at least 10 randomly selected locations on each sample, and the averaged spectra were utilized for subsequent analysis. The structural quality of the SWCNTs was assessed using the intensity ratio of the G-band
 (1590 cm\textsuperscript{-1}) to the D-band (1300-1350 cm\textsuperscript{-1}), denoted as
\textit{I}\textsubscript{G}/\textit{I}\textsubscript{D}.\cite{dresselhaus2005raman,saito2011raman}
The \textit{I}\textsubscript{G}/\textit{I}\textsubscript{D} ratio is indicative of the surface defect density of the SWCNTs.\cite{belin2005characterization}
The radial breathing modes (RBMs), ranging from 100 to 300 cm\textsuperscript{-1} in Raman spectra, were analyzed to determine the diameter \textit{d}\textsubscript{t} of SWCNTs, based on the
equation:\cite{bachilo2002structure}
$$\omega\textsubscript{RBM}(cm\textsuperscript{-1}) = 223.5 /\textit{d}\textsubscript{t} (nm) + 12.5$$

Morphological characterization of the SWCNTs was performed using SEM (S-4800, Hitachi) and TEM (JEM-2100 plus, JEOL) at acceleration voltages of 1 kV and 200 kV, respectively. SEM observations of the ND-CNTs was performed directly on the SiO\textsubscript{2}/Si substrates. For TEM analysis, the ND-CNTs were sonicated in ethanol for 1 h, and the resulting dispersion was drop-cast onto a copper TEM grid coated with a holey carbon film. As for SG-CNTs, powder was sonicated in ethanol for 30 min and the resultant solution was drop-cast onto a clean SiO\textsubscript{2}/Si substrate for SEM observations and a TEM grid for TEM analysis.

Thermogravimetric analysis (TGA) was performed using a TG8120 analyzer (Rigaku) on SG-CNTs samples weighing between 0.5 and 1.5 mg under a continuous flow of 200 sccm dry air. The samples were heated at a rate of 5°C/min from RT to 800°C. The peak temperature \textit{T}\textsubscript{max} derived from derivative thermogravimetry (DTG) was used as an indicator of the quality and defect density of the SWCNTs. A higher \textit{T}\textsubscript{max} suggests greater stability, higher quality, and fewer defects and
impurities.\cite{bom2002thermogravimetric,chiang2001purification}

\section{Results and discussion}
\subsection{Effect of multiple-cycle defect healing on ND-CNTs}
The Raman spectra of ND-CNTs before and after multiple-cycle defect healing with air exposure are presented in Figure 3(a). To examine the impact of the multiple-cycle defect-healing process, the \textit{I}\textsubscript{G}/\textit{I}\textsubscript{D} ratio of ND-CNTs was evaluated after each cycle. The normalized \textit{I}\textsubscript{G}/\textit{I}\textsubscript{D} ratio, based on the value from the pristine samples without healing, is presented as black square plots in Figure 3(b). Data were collected from at least 10 randomly selected Raman measurements, with error bars in Figure 3(b) representing the standard deviation (SD) of the normalized \textit{I}\textsubscript{G}/\textit{I}\textsubscript{D} across  different Raman datasets. Normalization was applied because the initial \textit{I}\textsubscript{G}/\textit{I}\textsubscript{D} of ND-CNTs exhibited variation across different growth batches. Using the normalized \textit{I}\textsubscript{G}/\textit{I}\textsubscript{D} ratio facilitates easier comparison of changes when the initial \textit{I}\textsubscript{G}/\textit{I}\textsubscript{D} ratios differ. The unnormalized \textit{I}\textsubscript{G}/\textit{I}\textsubscript{D} ratio of ND-CNTs is shown in Figure S2.

\begin{figure}[htbp]
  \centering
   \includegraphics[width=16cm]{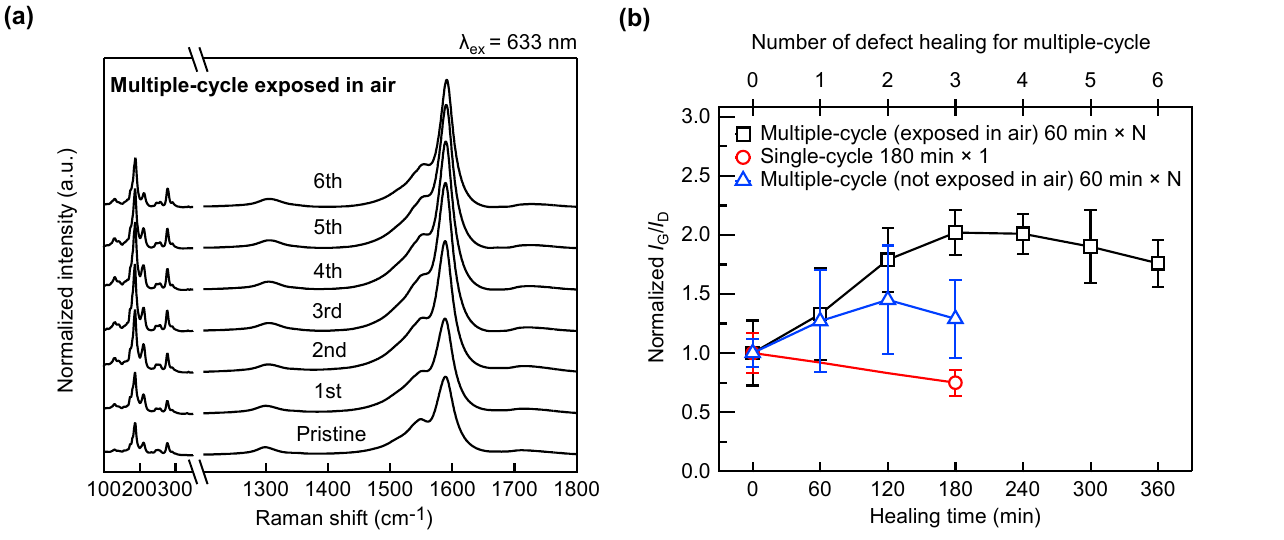}
    \caption{(a) Raman spectra of ND-CNTs before and after each cycle; (b) Changes in normalized \textit{I}\textsubscript{G}/\textit{I}\textsubscript{D} 
    with the increasing healing time during multiple-cycle defect healing (60 min × N, N = 0, 1, 2, 3…) with and without air exposure, as well as single-cycle defect healing (180 min × 1) of ND-CNTs.}\label{figure3}
\end{figure}

When the multiple-cycle defect-healing process was applied to ND-CNTs, as illustrated in Figure 3(b) and Figure S2, the normalized \textit{I}\textsubscript{G}/\textit{I}\textsubscript{D} ratio initially increased from 1.00 to 1.33, 1.79, and 2.02 over the first three cycles, indicating a consistent improvement in crystallinity. However, beyond the third cycle, the ratio gradually declines to 2.01, 1.90, and 1.76, suggesting a saturation or degradation effect. The normalized \textit{I}\textsubscript{G}/\textit{I}\textsubscript{D} ratio reached its maximum value of 2.02 after the third cycle, followed by a gradual decrease. This trend indicates a reduction in the surface defect density of ND-CNTs from their pristine state through the third healing cycle, confirming that ND-CNTs can undergo further defect healing beyond the initial process. These findings suggest that repeated healing cycles can be used to produce high-quality SWCNTs. It is also important to note that after the third healing cycle, the \textit{I}\textsubscript{G}/\textit{I}\textsubscript{D} ratio begins to decrease during subsequent cycles. This may be attributed to the deposition of amorphous carbon (a-C) caused by the introduction of C\textsubscript{2}H\textsubscript{2}\cite{jorio2003characterizing,wang2022thermal}.

To determine whether the healing effects observed in the multiple-cycle defect-healing process are simply due to prolonged annealing, ND-CNTs were subjected to a stepwise healing process in 5 sccm C\textsubscript{2}H\textsubscript{2} (2\%)/Ar for 180 min. As shown by the red circle plots in Figure 3(b) and Figure S2, continuous annealing of ND-CNTs under identical conditions resulted in a decrease in the normalized \textit{I}\textsubscript{G}/\textit{I}\textsubscript{D} ratio from 1.00 to 0.75. Despite the total healing time being identical, the single-cycle process produced significantly different results compared to the multiple-cycle process (black square plots in Figure 3(b)), where the normalized \textit{I}\textsubscript{G}/\textit{I}\textsubscript{D} ratio increased to 2.02. These findings indicate that the enhanced defect healing achieved through the multiple-cycle process cannot be attributed solely to extended heating durations. This observation aligns with our previous study, which demonstrated that suboptimal healing conditions over extended durations can lead to reductions in \textit{I}\textsubscript{G}/\textit{I}\textsubscript{D} values.\cite{wang2022thermal}

To investigate the multiple-cycle defect-healing process further, experiments were conducted on ND-CNTs subjected to multiple-cycle healing without air exposure (represented by the blue triangle plots in Figure 3(b) and Figure S2). The results show that even in the absence of air exposure, the normalized \textit{I}\textsubscript{G}/\textit{I}\textsubscript{D} ratio of ND-CNTs increased from 1.00 to 1.27 and 1.45 after the first and second healing cycles, respectively, confirming the inherent feature of the multiple-cycle process. Furthermore, the normalized \textit{I}\textsubscript{G}/\textit{I}\textsubscript{D} ratio after the first cycle was 1.33 under air-exposed conditions and 1.27 under non air-exposed conditions, demonstrating the consistency of the data despite different conditions. Notably, the normalized \textit{I}\textsubscript{G}/\textit{I}\textsubscript{D} ratio of the air-exposed ND-CNTs further increased to 1.79 and 2.02 after the second and third healing cycles, respectively. In contrast, the ratio for ND-CNTs not exposed to air increased to the maximum after the second cycle and subsequently decreased  after the third cycle. These results indicate that while the multiple-cycle defect-healing process is inherently effective, air exposure significantly enhances the healing efficiency for ND-CNTs. Thus, both the multiple-cycle process and air exposure are essential factors in promoting defect healing in ND-CNTs.

\subsection{Analysis of the D-band shape}
To investigate the factors contributing to the effectiveness of multiple-cycle healing and air exposure for defect healing in ND-CNTs, we analyzed the D-band shape, as shown in Figure 4. The analysis reveals that the D-band becomes broader after the healing process. Specifically, as illustrated in Figure 4(a), the full width at half maximum (FWHM) of the D-band increased from 43~cm$^{-1}$ before healing to 47~cm$^{-1}$ after air-exposed multiple-cycle healing. Furthermore, the FWHM increased to 57~cm$^{-1}$ following non air-exposed multiple-cycle healing and 59~cm$^{-1}$ after single-cycle healing. This broadening indicates the formation of a-C during the defect-healing process\cite{jorio2003characterizing}, primarily due to the introduction of C\textsubscript{2}H\textsubscript{2}. These findings are consistent with previous studies on SWCNTs, which also reported a-C formation during C\textsubscript{2}H\textsubscript{2}-assisted healing processes.\cite{wang2022thermal}  
Interestingly, the D-band broadens more significantly after a single 180-min healing cycle compared to multiple 180-min cycles with air exposure as illustrated in Figure 4(a). This suggests that a greater amount of a-C is deposited during the single-cycle process. In contrast, the multiple-cycle healing procaess reduces a-C deposition, thereby minimizing its impact on the D-band. This difference may be attributed to variations in defect activity between the single-cycle and multiple-cycle processes. Defect regions of ND-CNTs are more prone to adsorb gas molecules such as C\textsubscript{2}H\textsubscript{2}, and during the healing process, C\textsubscript{2}H\textsubscript{2} simultaneously heal the vacancies and decomposes to deposit a-C. For more active defects, leading to reduced C\textsubscript{2}H\textsubscript{2} absorption and, consequently, less a-C deposition. Furthermore, the D-band broadens more after no air exposure compared to air-exposed healing, as illustrated in Figure 4(a). This indicates that air exposure effectively reduces a-C deposition. Oxygen and other gas molecules in the air likely adsorb onto defect sites, inhibiting a-C formation at these locations, thereby limiting its accumulation. Additionally, at elevated temperatures, oxygen and other gas molecules may form functional groups at the defect sites. They may also induce mild etching effects. These processes activate the defects more effectively, making them more amenable to healing as indicated later in Figure 7.
Figure 4(b) further shows the FWHM of ND-CNTs before and after multiple-cycle defect healing without air exposure. Although the samples originated from different batches, their pristine conditions are considered equivalent. The FWHM increases from 45~cm$^{-1}$ to 48, 55, and 57~cm$^{-1}$ after the first, second, and third defect-healing cycles, respectively. This progressive increase in FWHM with additional cycles demonstrates the deposition of a-C during the multiple-cycle defect-healing process without air exposure. 

\begin{figure}[htbp]
  \centering
   \includegraphics[width=16cm]{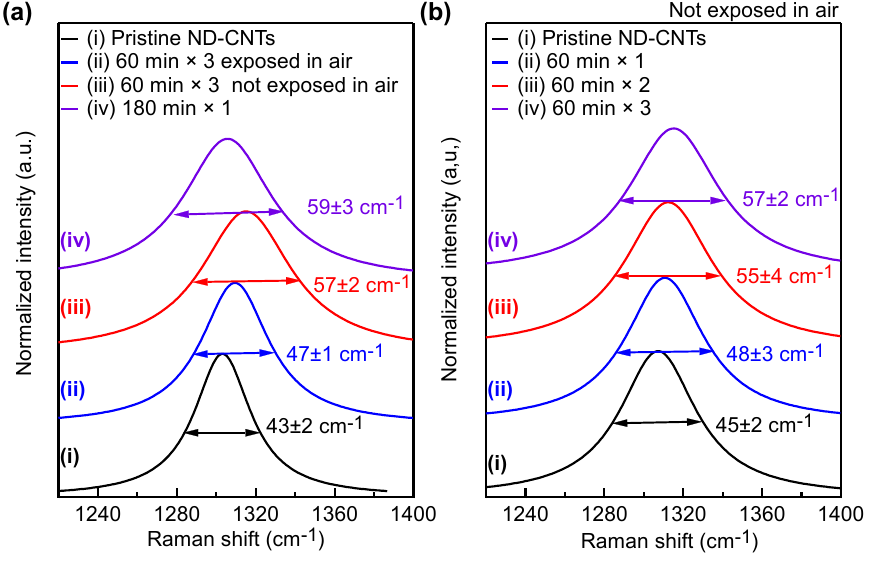}
    \caption{D-band  normalized  Raman  spectra  of ND-CNTs (a)  before  and  after  healed  with 
    C\textsubscript{2}H\textsubscript{2} through multiple-cycle (60 min × 3) with and  without air exposure as well as single-cycle (180 min × 1); (b) before and after healed with C\textsubscript{2}H\textsubscript{2} through multiple-cycle without air exposure.}
    \label{figure4}
\end{figure}

\subsection{Diameter distribution evaluated from the TEM images and Raman spectra}
To determine whether the diameter of ND-CNTs changes after multiple cycles of defect healing, we analyzed their diameters using Raman spectroscopy and TEM imaging. It is important to note that RBM peaks arise from resonance, meaning that only SWCNTs that meet the resonance conditions can generate these peaks. As a result, Raman spectroscopy cannot detect all SWCNTs, only SWCNTs with a transition energy of 1.96 eV (corresponding to 633 nm) were detectable in this study. The discussion here is limited to the SWCNTs detectable by Raman spectroscopy. As shown in Figure 3(a), the Raman spectra of ND-CNTs display five prominent RBM peaks at approximately 135, 190, 210, 250, and 280~cm$^{-1}$, corresponding to SWCNTs diameters of 1.8, 1.3, 1.1, 0.9, and 0.8~nm, respectively. While the intensities of the RBM peaks vary after the defect healing process, their RBM frequencies remain almost unchanged, indicating that the diameters of the ND-CNTs are preserved before and after defect healing. Additionally, TEM images were analyzed to evaluate the diameter distributions of ND-CNTs before and after each cycle of defect healing with air exposure. Figure 5 presents the diameter distributions of ND-CNTs observed via the TEM: (a) before healing, (b) after the first healing cycle, (c) after the second healing cycle, and (d) after the third healing cycle. The average diameters of ND-CNTs exhibited negligible changes, shifting from 1.18~nm to 1.22~nm, 1.13~nm, and 1.19~nm after the first, second, and third healing cycles, respectively. Raman spectroscopy and TEM analysis demonstrated that our defect-healing method at 1000°C effectively prevents changes in the diameters of the SWCNTs, in contrast to the traditional high-temperature methods at 1800~°C~\cite{metenier2002coalescence}, which caused a significant increase in the CNT diameters from 2 to 4~nm.

\begin{figure}[htbp]
  \centering
   \includegraphics[width=16cm]{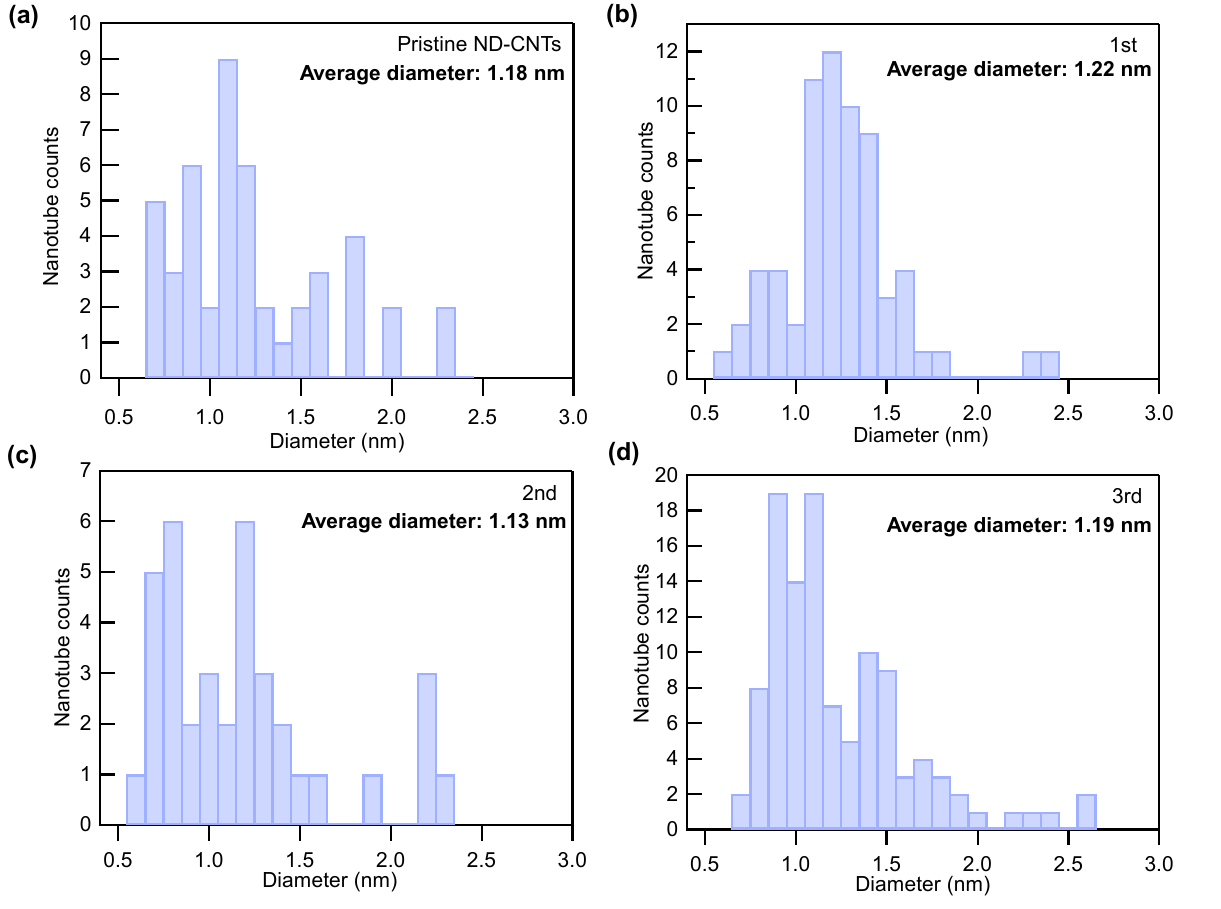}
    \caption{The diameter distributions of ND-CNTs analyzed using TEM images: (a) before defect healing (pristine) and, after the (b) 1st, (c) 2nd, and (d) 3rd defect healing cycle with air exposure.}
    \label{figure5}
\end{figure}

\subsection{Effect of multiple-cycle defect healing on SG-CNTs}
Multiple-cycle defect-healing process was also applied to SG-CNTs. Due to their larger quantity, SG-CNTs could be evaluated using TGA in addition to Raman spectroscopy. The TGA and DTG curves of SG-CNTs before healing, after heating in 25 sccm Ar, and following the first healing process of multiple-cycle with air exposure using \text{C}\textsubscript{2}\text{H}\textsubscript{2} are presented in Figure 6(a). The TGA and DTG curves after the second and third healing processes are almost identical, with minimal difference, as shown in Figure S3. The summarized \textit{I}\textsubscript{G}/\textit{I}\textsubscript{D} ratio and \textit{T}\textsubscript{max} values after each healing process are shown in Figure 6(b) (original Raman spectra are provided in Figure S4). Notably, the changes in \textit{T}\textsubscript{max} and \textit{I}\textsubscript{G}/\textit{I}\textsubscript{D} are consistent supporting the reliability of the results. \textit{T}\textsubscript{max} rose from 641.0°C to 650.5°C, indicating a reduction in the defect density and improvements in crystallinity and oxidation stability. These findings suggest an overall enhancement in the quality of the SG-CNTs. In addition, TEM images of SG-CNTs, shown in Figure S5, reveal that their diameters and structures remain almost unchanged (from 3.04~nm to 3.17~nm) after defect healing using C\textsubscript{2}H\textsubscript{2}.

When SG-CNTs were heated in an Ar atmosphere alone, as shown in Figure 6(a), the TGA curve shifted downward, indicating a decrease in quality. Conversely, when SG-CNTs were healed with \text{C}\textsubscript{2}\text{H}\textsubscript{2}, the TGA curve shifted upward, further emphasizing the critical role of \text{C}\textsubscript{2}\text{H}\textsubscript{2} in enhancing the quality of SG-CNTs. This observation is consistent with the case of ND-CNTs in our work, which demonstrated that \text{C}\textsubscript{2}\text{H}\textsubscript{2} improves the quality of SWCNTs.\cite{wang2022thermal}. However, after the first healing cycle, as shown in Figure 3(b), ND-CNTs continued to show improvement in subsequent cycles, whereas as shown in Figure 6(b) the \textit{I}\textsubscript{G}/\textit{I}\textsubscript{D} ratio and \textit{T}\textsubscript{max} of SG-CNTs remain relatively unchanged. This suggests that the defect density and crystallinity of SG-CNTs become stable after the first cycle. In addition, multiple-cycle defect healing without air exposure was conducted on SG-CNTs. Figures 6(c) and 6(d) presents the TGA curves and the summarized \textit{I}\textsubscript{G}/\textit{I}\textsubscript{D} ratio and \textit{T}\textsubscript{max} values after each healing process. By comparing Figures 6(b) and 6(d), we find that healing processes with and without air exposure yielded similar results for SG-CNTs, demonstrating that, unlike ND-CNTs, air exposure has minimal impact on the defect healing of SG-CNTs. 

The \textit{I}\textsubscript{G}/\textit{I}\textsubscript{D} ratio of SG-CNTs is lower than that of ND-CNTs, indicating a higher defect density in SG-CNTs. Although SG-CNTs are healed significantly after the first defect  cycle, there is little improvement in the \textit{I}\textsubscript{G}/\textit{I}\textsubscript{D} ratio after the second and third cycles. Even after healing, the \textit{I}\textsubscript{G}/\textit{I}\textsubscript{D} ratio of SG-CNTs remains lower than that of ND-CNTs, suggesting that the defect density in SG-CNTs remained higher than that in ND-CNTs. This higher defect density implies that defects in SG-CNTs persist through the second and third healing cycles, making it more difficult for them to heal. In other words, the healing reactivity of SG-CNTs is lower than that of ND-CNTs.

\begin{figure}[htbp]
  \centering
   \includegraphics[width=16cm]{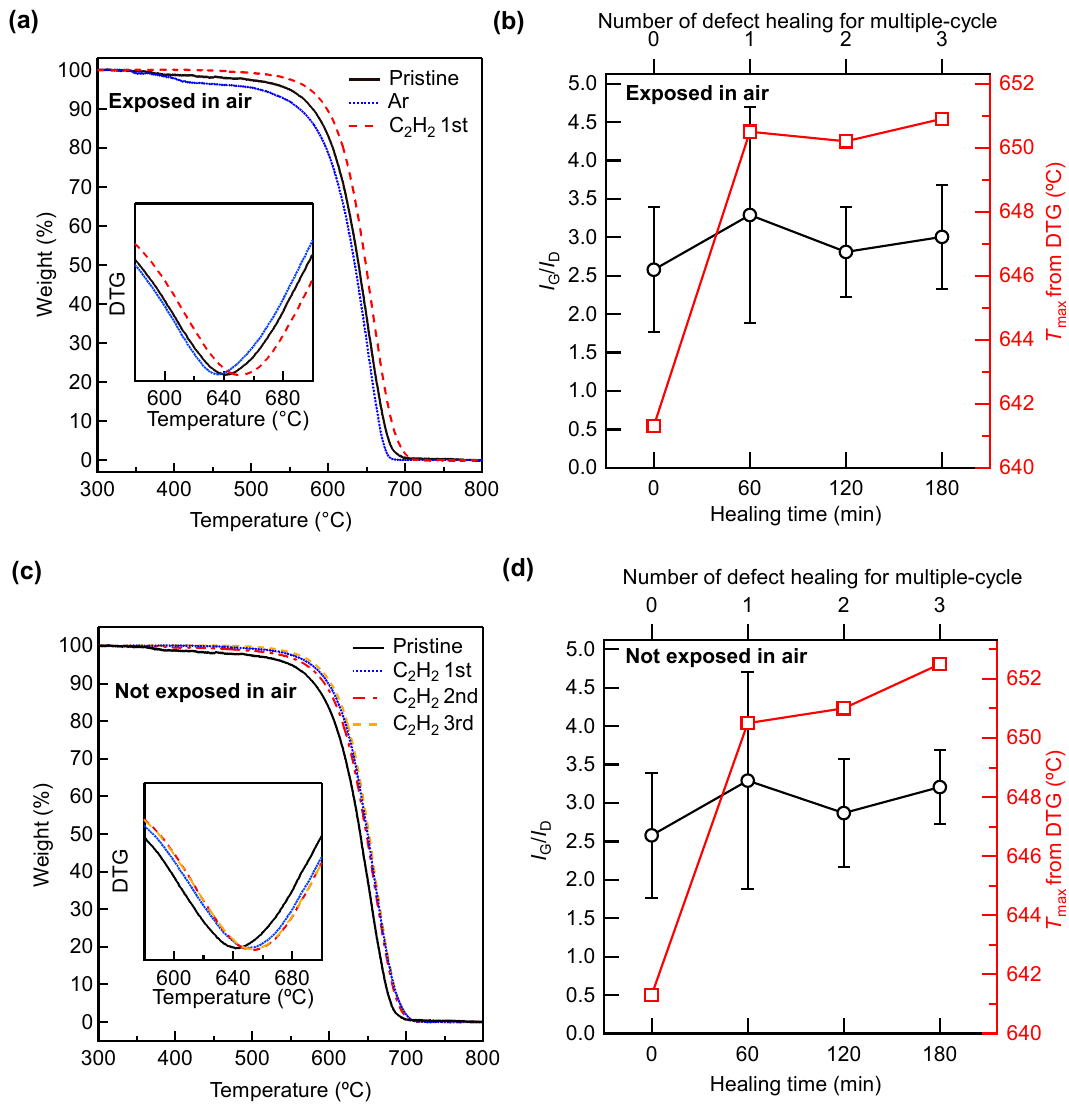}
    \caption{(a,c) TGA curves of SG-CNTs before and after multiple-cycle defect healing (a) with and (c) without air exposure; (b,d) Change of \textit{I}\textsubscript{G}/\textit{I}\textsubscript{D} and
    \textit{T}\textsubscript{max} of SG-CNTs during multiple-cycle defect healing (b) with and (d) without air exposure.}
    \label{figure6}
\end{figure}
This lower reactivity of SG-CNTs compared to ND-CNTs can be primarily attributed to differences in their diameters. The average diameter of ND-CNTs is approximately 1.2 nm, whereas SG-CNTs typically have diameters ranging from 3 to 5 nm. This difference in diameter significantly influences their structural stability and chemical reactivity.\cite{warner2009investigating} As shown in Figure S6, thinner CNTs, such as ND-CNTs, exhibit higher bond strain due to their larger curvature, resulting in lower stability but enhanced chemical reactivity. In contrast, SG-CNTs, with their smaller curvature, experience reduced bond strain, making them more stable but less reactive. This behavior is consistent with prior research showing that SWCNTs with smaller diameters (e.g., 0.8-1.1 nm) are more effectively healed by \text{C}\textsubscript{2}\text{H}\textsubscript{2} than those with larger diameters, (e.g., 1.3-1.8 nm) under similar conditions.\cite{wang2022thermal} Consequently, ND-CNTs undergo efficient healing across multiple cycles. The higher reactivity of the thinner ND-CNTs facilitates defect mobility and enhances the healing efficiency across multiple-cycles. Conversely, SG-CNTs exhibit limited healing efficiency, typically restricted to the first cycle, due to their reduced reactivity. These findings underscore the critical role of diameter in governing the defect-healing behavior of SWCNTs.  

\subsection{Model of multiple-cycle defect healing}
We summarize the healing process model developed in this study in Figure 7. As previously discussed, both multiple-cycle and air exposure play significant roles in reducing a-C deposition during defect healing. As shown in Figure 7(a) and (b), the differences in defect healing between the single-cycle and multiple-cycle processes (withourtair exposure) can be attributed to the variations in defect activity during each cycle. For ND-CNTs, defect sites are more likely to adsorb gas molecules like C\textsubscript{2}H\textsubscript{2}. During the healing process, C\textsubscript{2}H\textsubscript{2} not only fills vacancies but also decomposes and reacts with each other, resulting in a-C deposition. For highly active defects, the healing time is shorter, which reduces the adsorption duration of C\textsubscript{2}H\textsubscript{2} molecules on the surface, thereby minimizing a-C deposition. The repeated temperature cycling inherent in the multiple-cycle process enhances defect reactivity. Consequently, during the multiple-cycle process, defects are healed more efficiently, achieving a stable healing state in less time compared to single-cycle processing.

The impact of air exposure is illustrated in Figure 7(b) and Figure 7(c). Oxygen and other gas molecules in the air are more likely to adsorb onto defect sites, potentially inhibiting the formation of a-C at these locations and thereby reducing its accumulation. Additionally, at elevated temperatures, these gas molecules may form functional groups at defect sites or induce mild etching effects. These processes enhance the reactivity of the defects, making them more susceptible to healing.

\begin{figure}[htbp]
  \centering
   \includegraphics[width=16cm]{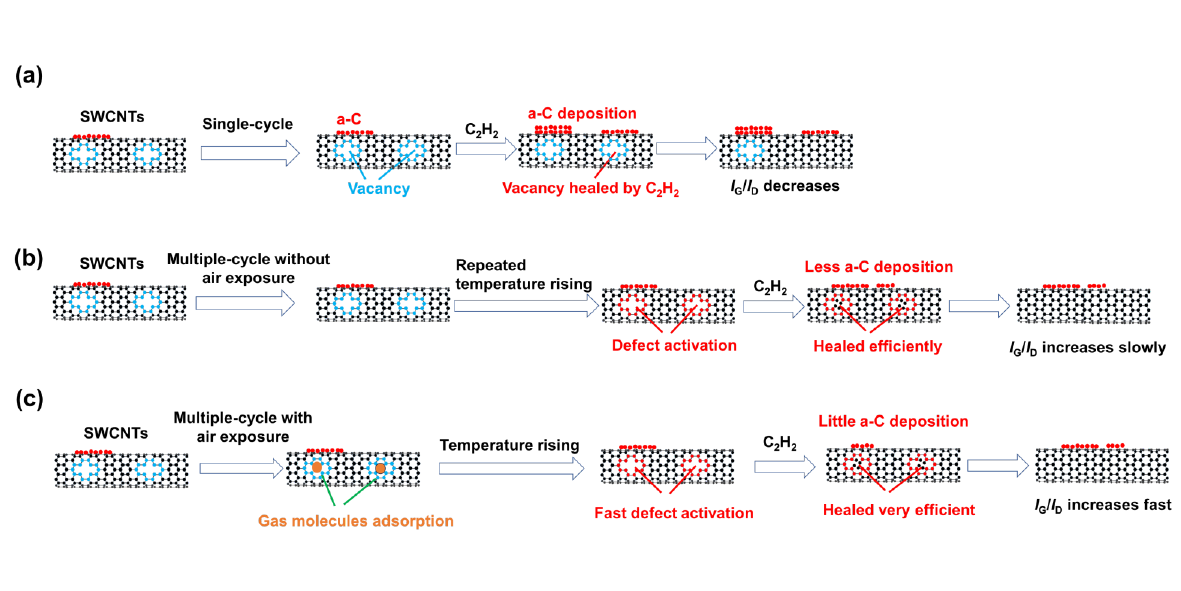}
    \caption{Model of SWCNTs during (a) single-cycle defect healing; (b)multiple-cycle defect healing without air exposure; (c) multiple-cycle defect healing with air exposure.}
    \label{figure7}
\end{figure}

\section{Conclusion}
In conclusion, we successfully achieved efficient defect healing of ND-CNTs through multiple-cycle defect-healing process, avoiding the structural changes typically induced by high-temperature treatments. The effects of multiple-cycle and single-cycle healing are distinctly different, largely due to the temperature cycling and air exposure involved in the multiple-cycle process. Air exposure plays a critical role by increasing defect activity through the adsorption of oxygen and other gas molecules at defect sites, inhibiting the formation of a-C, and promoting the generation of functional groups or mild etching effects at elevated temperatures. These mechanisms significantly accelerate defect healing while minimizing a-C deposition. In addition, the multiple-cycle defect-healing process was applied to commercially available SG-CNTs. The healing effect on SG-CNTs was primarily observed during the first cycle, with subsequent steps having minimal impact. This suggests that C\textsubscript{2}H\textsubscript{2} exhibits varying healing efficiencies across different types of SWCNTs, highlighting the necessity of tailoring the healing conditions based on the specific characteristics of the SWCNTs being used. The efficient healing with structural preservation demonstrated in this study is expected to be applicable to structurally sorted SWCNTs, enhancing their crystallinity and performance without compromising their inherent structure. This approach holds significant potential for applications requiring SWCNTs with high crystallinity and uniform chirality, such as solar cells, field-effect transistors, and quantum light sources.

\begin{acknowledgement}
The authors thanks M. Arifuku and N. Kiyoyanagi of the Nippon Kayaku for the supply of high purity nanodiamond. The authors also thanks Dr. T. Sakata and Dr. S. Takagi of the Research Center for Ultra-High Voltage Electron Microscopy, Osaka University, for assistance with the SEM and TEM observations. This research is supported by JSPS KAKENHI (JP15H05867, JP17H02745, and JP24K01297) and JST SPRING (JPMJSP2138).

\end{acknowledgement}
\providecommand{\latin}[1]{#1}
\makeatletter
\providecommand{\doi}
  {\begingroup\let\do\@makeother\dospecials
  \catcode`\{=1 \catcode`\}=2 \doi@aux}
\providecommand{\doi@aux}[1]{\endgroup\texttt{#1}}
\makeatother
\providecommand*\mcitethebibliography{\thebibliography}
\csname @ifundefined\endcsname{endmcitethebibliography}  {\let\endmcitethebibliography\endthebibliography}{}

\begin{suppinfo}
The following files are available free of charge.
\begin{itemize}
  \item Support information: The preparation of ND-CNTs, the \textit{I}\textsubscript{G}/\textit{I}\textsubscript{D} ratio of ND-CNTs during multiple-cycle defect healing, original Raman spectra of SG-CNTs during multiple-cycle defect healing with air exposure, TGA and DTG curves of SG-CNTs after second and third multiple-cycle defect healing with air exposure, TEM images of SG-CNTs before and after defect healing, Model of ND-CNTs and SG-CNTs during multiple-cycle defect healing.
\end{itemize}

\end{suppinfo}

\end{document}